\documentclass[showpacs,prd]{revtex4}
\usepackage{epsfig}
\usepackage{graphicx}
\usepackage{dcolumn}
\usepackage{amsmath}
\usepackage{latexsym}

\begin{document} 

\begin{flushright}
NITheP-09-07\\
\end{flushright}

\title{On the uniqueness of unitary representations of the non commutative Heisenberg-Weyl algebra}  
\date{\today}
\author{L Gouba $^a$\footnote{E-mail:
gouba@sun.ac.za} and F G Scholtz$^{a,b}$\footnote{E-mail: fgs@sun.ac.za}, }
\affiliation{$^a$National Institute for Theoretical Physics (NITheP), Stellenbosch Institute of Advanced Study, Stellenbosch 7600, South Africa\\
$^b$Institute of Theoretical Physics, University of Stellenbosch, Stellenbosch 7600, South Africa}

\begin{abstract}
In this paper we discuss the uniqueness of the unitary representations of the non commutative Heisenberg-Weyl algebra. We show that, apart from a critical line for the non commutative position and momentum parameters, the Stone-von Neumann theorem still holds, which implies uniqueness of the unitary representation of the Heisenberg-Weyl algebra.  
\end{abstract}

\maketitle

In commutative quantum mechanics a particle moving in $d$-dimensions is described by a configuration space $R^d$ and a Hilbert space $L^2$ of square integrable wave functions $\psi(x)$ over $R^d$.  A key element in the actual construction of a quantum system, as well as for the identification of the observables, is to find a unitary representation of the abstract Heisenberg algebra
\begin{eqnarray}
\label{heis}
\left[\hat{x}_i,\hat{p}_j\right]&=&i\hbar\delta_{i,j},\\\nonumber
\left[\hat{x}_i,\hat{x}_j\right]&=&0,\\
\left[\hat{p}_i,\hat{p}_j\right]&=&0,\nonumber
\end{eqnarray}
in terms of operators $\hat{x}_i$ and $\hat{p}_i$ acting on the space of square integrable functions.  This is just the well known Schr\"odinger representation
\begin{equation}
\label{sch}
\hat x_i\psi(x)=x_i\psi(x),\quad \hat p_i\psi(x)=-i\hbar\frac{\partial\psi(x)}{\partial x_i},
\end{equation}
which acts irreducibly and, from the Stone- von Neumann theorem, is known to be unique up to unitary transformations.
     
In the non commutative case the non commutative configuration space, which we take to be two dimensional for simplicity, is defined by the commutation relations
\begin{equation}
[\hat{x}_i, \hat{x}_j] = i\theta\epsilon_{i,j},
\end{equation}
where we take without loss of generality $\theta$ to be a real positive number, while $\epsilon_{i,j}$ is the completely anti-symmetric tensor.  Introducing the creation and annihilation operators 
\begin{eqnarray}\nonumber
b &=& \frac{1}{\sqrt{2\theta}} (\hat{x}_1+i\hat{x}_2),\\
b^\dagger &=&\frac{1}{\sqrt{2\theta}} (\hat{x}_1-i\hat{x}_2),
\end{eqnarray}
which satisfy the Fock algebra $[ b ,b^\dagger ] = 1$, this configuration space is isomorphic to boson Fock space
\begin{eqnarray}
\mathcal{H}_c = \textrm{span}\{ |n\rangle\equiv \frac{1}{\sqrt{n!}}(b^\dagger)^n |0\rangle\}_{n=0}^{n=\infty},
\end{eqnarray}
where the span is taken over the field of complex numbers. 

The Hilbert space in which the physical states of the system are to be represented, is the space of Hilbert-Schmidt operators acting on non commutative configuration space \cite{sch}
\begin{equation}
\label{qhil}      
\mathcal{H}_q = \left\{ \psi(\hat{x}_1,\hat{x}_2): \psi(\hat{x}_1,\hat{x}_2)\in \mathcal{B}\left(\mathcal{H}_c\right),\;{\rm tr_c}(\psi(\hat{x}_1,\hat{x}_2)^\dagger\psi (\hat{x}_1,\hat{x}_2)) < \infty \right\}.
\end{equation}
Here ${\rm tr_c}$ denotes the trace over non-commutative configuration space, $\mathcal{B}\left(\mathcal{H}_c\right)$ the set of bounded operators on $\mathcal{H}_c$ and the inner product is given by
\begin{equation}\label{inner}
\left(\phi(\hat{x}_1,\hat{x}_2),\psi(\hat{x}_1,\hat{x}_2)\right) = {\rm tr_c}(\phi(\hat{x}_1,\hat{x}_2)^\dagger\psi(\hat{x}_1,\hat{x}_2)).
\end{equation}

The abstract Heisenberg algebra is replaced by the non commutative Heisenberg algebra, which reads in two dimensions
\begin{eqnarray}
\label{heisnc}
\left[{\hat x}_i,{\hat p}_j\right] &=& i\hbar\delta_{i,j},\nonumber\\
\left[{\hat x}_i,{\hat x}_j\right] &=& i\theta\epsilon_{i,j},\\
\left[{\hat p}_i,{\hat p}_j\right] &=& 0\nonumber.
\end{eqnarray}
A unitary representation of this algebra in terms of operators $\hat{X}_i$ and $\hat{P}_i$, acting on $\mathcal{H}_q$, and which is the analog of the Schr\"{o}dinger representation of the Heisenberg algebra is given by \cite{sch} 
\begin{eqnarray}
\label{schnc}
\hat{X}_i\psi(\hat{x}_1,\hat{x}_2) &=& \hat{x}_i\psi(\hat{x}_1,\hat{x}_2),\nonumber\\
\hat{P}_i\psi(\hat{x}_1,\hat{x}_2) &=& \frac{\hbar}{\theta}\epsilon_{i,j}[\hat{x}_j,\psi(\hat{x}_1,\hat{x}_2)],
\end{eqnarray}
i.e., the position acts by left multiplication and the momentum adjointly.

The question that naturally arises at this point is whether this representation is unique, up to unitary transformations, as is the case with the Schr\"{o}dinger representation.  This is equivalent to the question whether the Stone-von Neumann theorem also applies in this case.  We address this issue next in a slightly more general setting where we also allow the momenta to be non commuting.  Note, however, that the representation (\ref{schnc}) only applies in the case of commuting momenta.

Let us consider the non commutative Heisenberg algebra 
\begin{eqnarray}
\left[\hat{x}_i,\hat{x}_j\right] &=& i\theta \epsilon_{i,j},\nonumber\\
\left[\hat{x}_i,\hat{p}_j\right] &=& i\hbar\delta_{i,j},\\
\left[\hat{p}_i,\hat{p}_j \right] &=& i\gamma \epsilon_{i,j},\nonumber
\end{eqnarray}
where $i,j=1,2$.  

The approach we shall take to proof the uniqueness of the unitary representations of these algebras, is to first find real linear combinations of the non commuting position and momenta that satisfy the standard Heisenberg algebra.  Once this has been achieved, we know from the Stone-von Neumann theorem that unitary representations of these linear combinations are equivalent and thus unique.  The unitary representations of the original algebra that can be derived from these must therefore also be unique up to unitary transformations.  For this argument it is essential to take real linear combination, otherwise the derived representations may not be unitary.  Conversely, if we can find a unitary representation of the original algebra, we can construct a unitary representation of the Heisenberg algebra through these linear combinations.  Again the uniqueness of the representations of the Heisenberg algebra imply that the unitary representations of the original algebra must be unique.  

We therefore seek real linear combinations of the coordinates and momenta that satisfy the Heisenberg commutation relations.  Let us set
\begin{eqnarray}
\label{posdel}
\hat{y}_i &=& \hat{x}_i + a\epsilon_{i,j}\hat{p}_j,\nonumber\\
\hat{q}_i &=& \hat{p}_i - b\epsilon_{i,j}\hat{x}_j,
\end{eqnarray}
with $a,b\in \Re$ and require them to satisfy the Heisenberg algebra
\begin{eqnarray}\label{cha}
\left[\hat{y}_i\:,\:\hat{y}_j \right] &=& 0,\nonumber\\
\left[\hat{q}_i\:,\:\hat{q}_j \right] &=& 0,\nonumber\\
\left[\hat{y}_i\:,\:\hat{q}_j \right] &=& i\Sigma\delta_{i,j}.
\end{eqnarray}

A simple calculation yields the following solutions for $a,b$:
\begin{eqnarray}
a^\pm = \frac{\hbar\pm\sqrt{\Delta}}{\gamma}\,,\quad
b^\pm = \frac{\hbar\pm\sqrt{\Delta}}{\theta}\,,\quad\Delta=\hbar^2 -\gamma\theta,
\end{eqnarray}
and the commutator
\begin{equation}
\left[\hat{y}_i\:,\:\hat{q}_j \right] =i\left(\hbar (1+ab) -(b\theta +\gamma a)\right)\delta_{i,j}\equiv i\Sigma\delta_{i,j}
\end{equation}
Since $\Sigma$ vanishes for the values $ (a,b) = (a^+, b^-)$ or $(a,b) = (a^-,b^+)$, the possibles transformations are
\begin{eqnarray}
\hat{y}^\pm_i &=&  \hat{x}_i +\frac{\hbar \pm \sqrt{\Delta}}{\gamma}\epsilon_{i,j}\hat{p}_j,\nonumber\\\label{lcomb}
\hat{q}^\pm_i &=& \hat{p}_i -\frac{\hbar \pm \sqrt{\Delta}}{\theta}\epsilon_{i,j}\hat{x}_j,
\end{eqnarray}
with the algebra
\begin{eqnarray}
\label{posdelheis}
\left[\hat{y}^\pm_i,\hat{y}^\pm_j \right] &=& 0,\nonumber\\
\left[\hat{q}^\pm_i, \hat{q}^\pm_j\right] &=& 0,\nonumber\\
\left[\hat{y}^\pm_i,\hat{q}^\pm_j\right] &=& 2i\frac{\Delta}{\gamma\theta}(\hbar \pm \sqrt{\Delta})\delta_{i,j}.
\end{eqnarray}
In the case of $\Delta > 0$, this algebra is completely isomorphic to the Heisenberg algebra, which is obtained by a simple scaling transformation that replaces $2\frac{\Delta}{\gamma\theta}(\hbar \pm \sqrt{\Delta})$ by $\hbar$.  Furthermore the linear combinations in (\ref{lcomb}) are then real.   

In the case of $\Delta < 0$ new coordinates and momenta are constructed as follows 
\begin{eqnarray}
\label{negdel}
\hat{y}_1 = \gamma\hat{x}_2 + a\hat{p}_1,\nonumber\\
\hat{y}_2 =  \theta\hat{p}_2  -a\hat{x}_1,\nonumber \\
\hat{q}_1 = b\hat{x}_1 -\theta\hat{p}_2,\nonumber\\
\hat{q}_2 =\gamma\hat{x}_2 + b\hat{p}_1.
\end{eqnarray}
Requiring them to satisfy the Heisenberg algebra yields the following solutions for $a,b$
\begin{eqnarray}
a^\pm  = \frac{-\gamma\theta \pm \sqrt{-\gamma\theta\Delta}}{\hbar},
\:\:\: \:
b^\pm  = \frac{-\gamma\theta \pm \sqrt{-\gamma\theta\Delta}}{\hbar},
\end{eqnarray}
as well as the commutator
\begin{eqnarray}
[\hat{y}_i,\hat{q}_j] = -i\left(\hbar(\gamma\theta+ab)+\gamma\theta(a+b)
\right)\delta_{i,j} \equiv i\Sigma\delta_{i,j}.
\end{eqnarray}
For the values $(a,b) = (a^+,b^+)$ or $(a,b) = (a^-,b^-)$, $\Sigma=0$ so that the possible transformations are
\begin{eqnarray}
\hat{y}_1^\pm &=& \gamma \hat{x}_2 + \frac{-\gamma\theta \pm 
\sqrt{-\gamma\theta\Delta}}{\hbar}\hat{p}_1,\nonumber\\
\hat{y}_2^\pm &=& \theta \hat{p}_2 -
 \frac{-\gamma\theta \pm\sqrt{-\gamma\theta\Delta}}{\hbar}\hat{x}_1,\nonumber \\
\hat{q}_1^\mp &=& -\theta\hat{p}_2+ \frac{-\gamma\theta\mp \sqrt{-\gamma\theta\Delta}}{\hbar}\hat{x}_1, \nonumber\\
\hat{q}_2^\mp &=& \gamma\hat{x}_2+\frac{-\gamma\theta \mp \sqrt{-\gamma\theta\Delta}}{\hbar}\hat{p}_1,
\end{eqnarray}
with algebra 
\begin{eqnarray}
\label{negdelheis}
\left[\hat{y}^\pm_i,\hat{y}^\pm_j \right] &=& 0,\nonumber\\
\left[\hat{y}^\pm_i,\hat{q}^\mp_j\right] &=& 
-2i\frac{\gamma\theta}{\hbar}\Delta\delta_{i,j},\nonumber\\
\left[\hat{q}^\mp_i, \hat{q}^\mp_j\right] &=& 0.
\end{eqnarray}
For $\Delta < 0$, this algebra is again completely isomorphic to the Heisenberg algebra, which is obtained by a simple scaling transformation that replaces $-2\frac{\gamma\theta}{\hbar}\Delta$ by $\hbar$ and the linear combinations in (\ref{negdel}) are real.   

If we have a set of hermitian operators $\hat{y}^\pm_i$, $\hat{q}_j^\pm$ representing the Heisenberg algebras (\ref{posdelheis}) and (\ref{negdelheis}) on some Hilbert space, we can construct a set of hermitian operators satisfying the original non commutative Heisenberg algebra via (\ref{posdel}) and (\ref{negdel}), respectively. In this regard the distinction between $\Delta > 0$ and  $\Delta<0$ is important to ensure hermiticity of these operators. Now we can also address the uniqueness of this representation by implementing the Stone-von Neumann theorem.  Associated with the hermitian operators $\hat{y}_i^\pm$, $\hat{q}_j^\pm$ that represent the Heisenberg algebras (\ref{posdelheis}) and (\ref{negdelheis}), we can construct from Stone's theorem\cite{holevo} a set of unitary bounded Weyl operators 
\begin{eqnarray}
U^\pm(\alpha) &=& e^{i\sum_{j=1}^2\alpha_j\hat{y}_j^{\pm}},\nonumber\\
V^\pm(\beta)  &=&  e^{i\sum_{j=1}^2\beta_j\hat{q}_j^\pm},
\end{eqnarray}
with $\alpha =(\alpha_1,\alpha_2),\beta = (\beta_1,\beta_2)\in R^2$.  These operators satisfy in the case of (\ref{posdelheis})
\begin{eqnarray}\nonumber
U^\pm (\alpha)U^\pm(\alpha') &=&  U^\pm(\alpha +\alpha'),\\\nonumber
V^\pm(\beta) V^\pm (\beta')  &=& V^\pm (\beta +\beta'),\\\label{sweyl1}
U^\pm(\alpha) V^\pm(\beta) &=& e^{i\omega^{\pm}_1(\alpha,\beta)}V^\pm(\beta) U^\pm(\alpha),
\end{eqnarray}
with
\begin{eqnarray}
\omega^\pm_1(\alpha,\beta) = -2\frac{\Delta}{\gamma\theta}(\hbar \pm \sqrt{\Delta}) (\alpha_1\beta_1 +\alpha_2\beta_2).
\end{eqnarray}
In the case of (\ref{negdelheis}) they satisfy
\begin{eqnarray}\nonumber
U^\pm (\alpha)U^\pm(\alpha') &=&  U^\pm(\alpha +\alpha'),\\\nonumber
V^\mp(\beta) V^\mp (\beta')  &=& V^\mp (\beta +\beta'),\\\label{sweyl2}
U^\pm(\alpha) V^\mp(\beta) &=& e^{i\omega_2(\alpha,\beta)}V^\mp(\beta) U^\pm(\alpha),
\end{eqnarray}
where
\begin{eqnarray}
\omega^{+}_2(\alpha,\beta) =\omega^{-}_2(\alpha,\beta)\equiv\omega_2(\alpha,\beta)= 2\Delta\frac{\gamma\theta}{\hbar}(\alpha_1\beta_1 +\alpha_2\beta_2).
\end{eqnarray}
For $\hbar^2 \neq \gamma\theta$, the forms $\omega_1^\pm, \:\omega_2$
are bilinear and non degenerate on $R^2$ and the Weyl systems (\ref{sweyl1})
and (\ref{sweyl2}) satisfy the Stone-von Neumann uniqueness theorem stated as in e.g. \cite{jona}: {\it  
Consider pairs (U,V) of unitary operators on a Hilbert space $\mathcal{H}$, satisfying the commutation rule
\begin{eqnarray}
U(x)V(y) = \exp(i\omega(x,y))V(y)U(x)
\end{eqnarray}
where $\omega: R^n\times R^n \rightarrow R$ is bilinear and non degenerate. Such pairs
are all equivalent to multiples of the standard Schr\"{o}dinger representation
on $L^2(R^n)$}.

This implies that the unitary representation $\hat{y}_i^\pm$, $\hat{q}_j^\pm$ of the Heisenberg algebras (\ref{posdelheis}) and (\ref{negdelheis}) is unique, up to unitary transformations, and therefore also the unitary representation $\hat{x}_i$, $\hat{p}_j$ of the original non commutative Heisenberg algebra constructed via (\ref{posdel}) and (\ref{negdel}).  As (\ref{schnc}) is a special case with $\gamma=0$, it follows that this representation is unique up to unitary transformations.  Note that in this case the linear combination with a smooth limit as $\gamma\rightarrow 0$ must be taken. 

The conclusions above, of course, fail on the critical line $\hbar^2 =\gamma\theta$ where the non degeneracy of the forms $\omega^\pm_1,\omega_2$ is lost and the Stone-von Neumann theorem does not apply.
Allowing the momenta to be non commuting entails two phases separated 
by the line $\hbar^2 = \gamma\theta$ where the Stone-von Neumann theorem does
hold. The two phases of the noncommutative quantum mechanics is 
also discussed in detail in \cite{stef}

\noindent{\bf Acknowledgements.}

This work was supported under a grant of the  National Research Foundation of South Africa. 


\end{document}